\documentclass[twocolumn,showpacs,preprintnumbers,amsmath,amssymb]{revtex4}
\usepackage{graphicx}
\usepackage{dcolumn}
\usepackage{bm}

\begin{document}


\title{Entanglement and quantum phase transitions}
\author{Shi-Jian Gu$^{1}$}\author{Guang-Shan Tian$^{1,2}$}
\author{Hai-Qing Lin$^{1}$}

\affiliation{$^1$Department of Physics and Institute of Theoretical
Physics, The Chinese University of Hong Kong, Hong Kong, China}
\affiliation{$^2$School of Physics, Peking University, Beijing 100871,
China}

\date{\today}

\begin{abstract}
We examine several well known quantum spin models and categorize behavior
of pairwise entanglement at quantum phase transitions. A unified picture on
the connection between the entanglement and quantum phase transition is
given.
\end{abstract}

\maketitle

Quantum phase transitions (QPTs) at zero temperature are characterized by
the change in the properties of the ground state of a many-body system
caused by modifications in the interactions among its constituents
\cite{Sachdev}. QPTs are induced as a parameter $g$ in the system
Hamiltonian $H(g)$ is varied across a point $g_c$. These phase transitions
are completely driven by the quantum fluctuations and are incarnated via
the non-analytic behaviors of the ground-state properties at the transition
points. On the other hand, as it is well known, the concept of entanglement
lies at the heart of quantum mechanics \cite{Einstein,Schrodinger}.
Therefore, one expects that quantum entanglement (QE) should play an
important role in QPTs. Recently, a great deal of effort has been devoted
to understanding their connection. Indeed, it has been observed that the
quantum phase transitions are signaled by critical behaviors of the
concurrence, a measurement of bipartite entanglement \cite{WKWootters98},
in a number of spin models
\cite{AOsterloh2002,TJOsbornee,GVidal2003,SJGu03,OFSyljuasen03,SJGu05,JVidal04,JVidal042,LAWu04,MFYang05,SJGu04J1J2}.
{\it However, they are not universal}. For example, for the
transverse-field Ising model, Osterloh {\it et. al.} found that the first
order derivative of the concurrence diverges at the transition point and
obeys a scaling law in its vicinity \cite{AOsterloh2002,TJOsbornee}. On the
other hand, for the antiferromagnetic $XXZ$ chain (1D), the concurrence
behaves in a completely different way. It is a continuous function of the
anisotropic parameter and reaches its maximum at the transition point
\cite{SJGu03}. While in two- and three-dimensional (2D \& 3D) $XXZ$
models\cite{SJGu05}, it develops a cusp-like behavior around the critical
point. Another interesting example is the so-called $J_1-J_2$ model (or
Majumdar-Ghosh model) \cite{CKMajumdar69}. Its ground state undergoes a
first-order phase transition at $J_2/J_1=0.5$ and a continuous phase
transition at $J_2/J_1\simeq 0.241$. However, unlike the previous cases,
the concurrence itself is now discontinuous at 0.5, while at 0.241, no
discernable structure has been found \cite{SJGu04J1J2}. Therefore, a
natural question arose is why the same quantity, which measures
entanglement between two localized spins, has such different behaviors,
such as singularity\cite{LAWu04}, maximum, scaling, etc., at QPTs? Further,
is there a unified picture of QE at the QPT? Obviously, investigation on
these issues will not only deepen our understanding in QPTs but also
strengthen the connection between condensed matter physics and quantum
information theory \cite{MANielsenb}.

In this Letter, we study these issues by detail analysis of several well
known spin models aim at giving a unified picture of QE at QPTs. For
definiteness, we choose the concurrence, $C$, as the measure of pairwise
entanglement\cite{WKWootters98} in this work. We characterize the behaviors
of the concurrence into \emph{three types} and emphasize the important role
played by the low-lying excitation spectra reconstruction of many-body
systems around the transition points in determining the critical behaviors
of the concurrence. More precisely, we show that the low-lying excitation
spectra of these models are reconstructed in three qualitatively different
ways around $g_c$ and hence, their concurrences show the above-mentioned
non-universal behaviors at the QPT points. Therefore, we are able to give a
unified and intuitive picture to understand the relation between the QPT
and the entanglement.

\begin{table*}
\caption{The basic features of typical spin models, such as the properties of
the concurrence, level-crossing(LC) in the ground-state (GS) and the first
excited-state(ES), symmetry at the transition point, and type of phase
transition. }
\begin{tabular}{|c|c|c|c|c|c|c|}
\hline\hline
Model (QPT point)  & GS LC & ES LC & concurrence & symmetry & transition type&
type

\\ \hline $XXZ$ chain($\Delta=-1$) & Yes
& & singular & & & I

\\ \hline $J_1-J_2$ model($J_2=0.5$) & Yes
& & singular & & & I

\\ \hline $XXZ$
chain($\Delta=1$)  & No & Yes &maximum, not singular & SU(2) point &
order-to-order & II

\\ \hline spin ladder($J=0$) & No & Yes & maximum, not singular & SU(2)$\otimes$SU(2)
 & disorder-to-disorder & II

\\ \hline $XXZ$ 2\&3D($\Delta=1$) & No & Yes& maximum,
singular & SU(2)  & order-to-order & II

\\ \hline $J_1-J_2$ model($J_2\simeq 0.241$) & No & Yes & not maximum & unknown  &
order-to-disorder & III

\\ \hline Ising model($\lambda=1$)  & No & No & singular, not maximum & unknown
 & disorder-to-order & III

\\ \hline\hline
\end{tabular}
\end{table*}

\begin{figure}
\includegraphics[width=7cm]{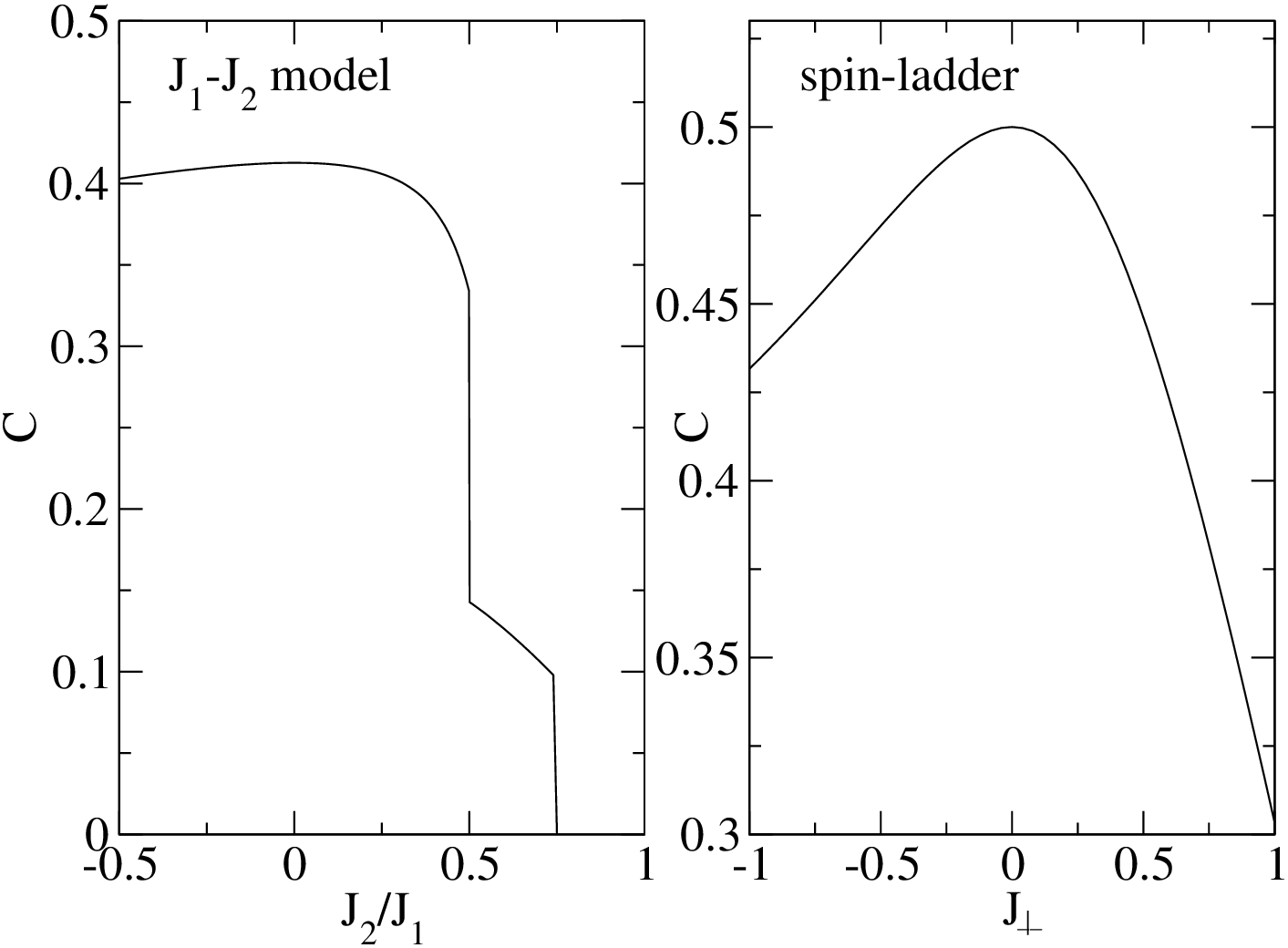}
\caption{\label{figure_ladder_fheis} Concurrence of a 8-site $J_1-J_2$
model (LEFT) and a $2\times 4$ spin-ladder system (RIGHT).\\}
\end{figure}

\begin{figure}
\includegraphics[width=7cm]{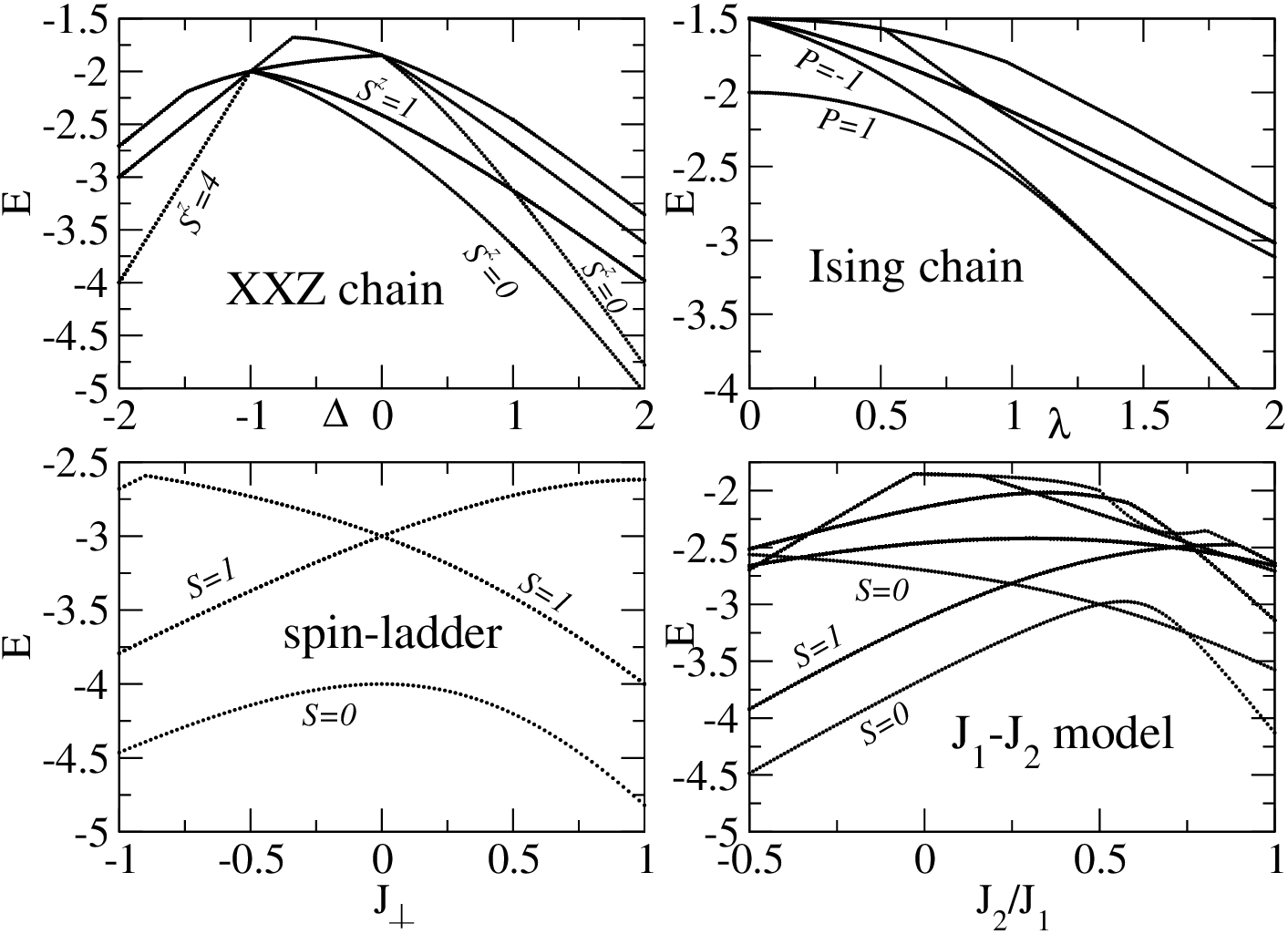}
\caption{\label{figure_energy} Low-lying energy spectra of four typical
spin models.}
\end{figure}

To have concrete discussions, we concentrate on three well studied quantum
spin models in one dimension, the $J_1-J_2$ model, the $XXZ$ model, and the
transverse-field Ising model, defined by the following Hamiltonians:
\begin{eqnarray}
\hat{H}_{\rm J_1-J_2} & = & \sum_i \left( J_1 \hat{\bf s}_i\cdot \hat{\bf
s}_{i+1} + J_2 \hat{\bf s}_i\cdot\hat{\bf s}_{i+2} \right); \\
\hat{H}_{\rm XXZ} & = & \sum_i \left( \hat{s}_i^x \hat{s}_{i+1}^x +
\hat{s}_i^y\hat{s}_{i+1}^y
+ \Delta \hat{s}_i^z \hat{s}_{i+1}^z \right); \\
\hat{H}_{\rm Ising} & = & - \sum_i \left( \lambda
\hat{s}_i^x\hat{s}_{i+1}^x + \hat{s}_i^z/2 \right).
\end{eqnarray}
In these Hamiltonians, $\hat{s}_i^x,\>\hat{s}_i^y$ and $\hat{s}_i^z$ denote
the spin operators at lattice site $i$. $J_1,\>J_2,\>\Delta$ and
$\lambda>0$ are interaction parameters. The periodic boundary condition
$\hat{s}_1 = \hat{s}_{N+1}$ is assumed. In Table I, we present basic
features of these three models. The concurrence of the $J_1-J_2$ model and
the spin ladder is exemplified in Fig. \ref{figure_ladder_fheis}, while the
concurrence of the Ising model and the $XXZ$ model can be found in Refs.
\cite{AOsterloh2002,TJOsbornee} and Refs.
\cite{SJGu03,OFSyljuasen03,SJGu05} respectively. As summarized in Table I,
there exist {\it three types} of QE at QPT points: (I) $C$ is
discontinuous; (II) $C$ is continuous and exhibits maximum at QPTs; (III)
$C$ is continuous at QPTs but its higher order derivative exhibits
extremum. We shall show that {\it these seemingly different critical
behaviors of the concurrence can be understood on the basis of the
low-lying spectra reconstruction of these models around their QPT points}.
For this purpose, we present the ground-state energy and some low
excited-state energies of the $J_1-J_2$ model, the $XXZ$ model, the spin
ladder model and the transverse-field Ising model, on a {\it finite} chain
in Fig. \ref{figure_energy}.

For type I, we see immediately that, in both the $J_1-J_2$ and the $XXZ$
models, there exist ground-state level-crossing at $J_2/J_1=0.5$ and
$\Delta=-1$, respectively. Since the concurrence is a measurement of
bipartite entanglement in the ground state, it is not difficult to see why
it changes discontinuously at these transition points. The same behavior
was also observed in the first order transition in the spin model with
mutual exchanges \cite{JVidal04}.

\begin{figure}
\includegraphics[width=7cm]{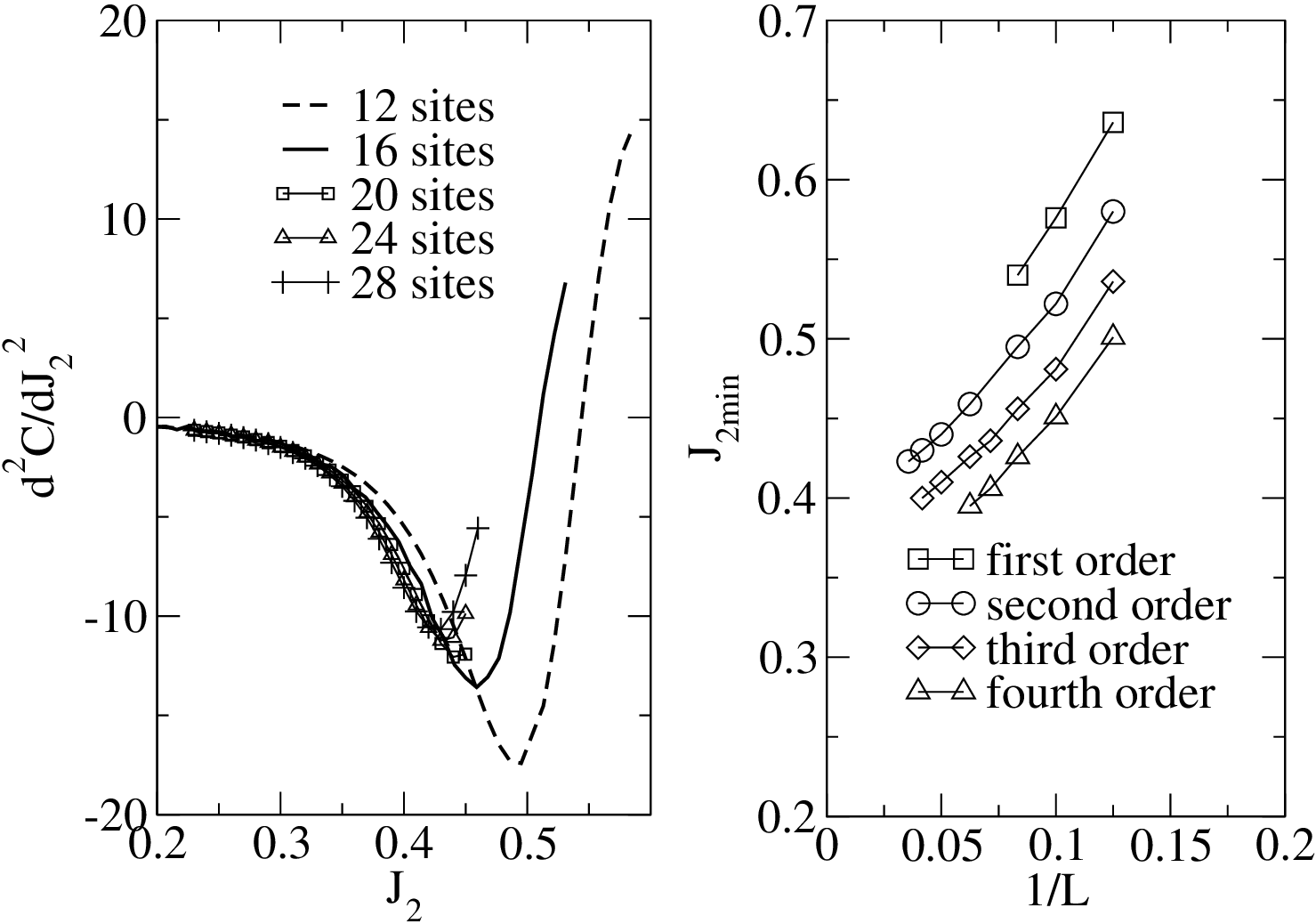}
\caption{\label{figure_j1j2dcon} LEFT: The second derivative of the
concurrence with respect to $J_2$ for various-size systems. RIGHT: The
scale of the minimum point of the derivatives of the concurrence with
different order.}
\end{figure}

To differentiate type II and III, we must take low-lying excitations into
account. Let us study the spectra of the $XXZ$ model at $\Delta=1$, the
spin ladder model at $J_\perp=0$, and the transverse-field Ising model at
$\lambda=1$. It has been proven that the ground states of these models are
nondegenerate around the corresponding transition points \cite{Lieb}. At
the first glance, it seems that this observation does not help us very much
in understanding the critical behaviors of the concurrence in these models.
However, as it is well known, QPTs are not solely dictated by the ground
state of a specific model, especially on a finite system. They also depend
on interconnection between the ground state and the low-lying excited
states of the system \cite{Sachdev}. In other words, the change of the
ground-state properties is greatly affected by matrix elements of the
relevant collective-mode operators, such as the spin-wave operators or the
particle-density-wave operators, which relates the low-lying excited states
to the ground state. In particular, in Ref. \cite{Tian03}, we pointed out
that the QPT is actually induced by the reconstruction of the excitation
energy spectra.
As a consequence, when a transition takes place between two ordered phases
of a system, such as the one for the $XXZ$ model at $\Delta=1$, which
separates the $XY$ regime from the Ising regime, a level-crossing in its
excited spectrum must occur. In fact, in Fig. \ref{figure_energy}, we see
clearly the existence of such a level-crossing between the first and second
excited states at $\Delta=1$ for the $XXZ$ model and at $J_\perp=0$ for the
spin ladder model. Both excited states have total spin $S=1$ and are
connected to the ground state, which is a spin singlet, via the spin-wave
operators. It also implies that the system enjoys higher symmetries at the
transition points. For instance, the $XXZ$ model has a $SU(2)$ symmetry at
$\Delta=1$ while its symmetry group is weakened to $SU_q(2)$ for $\Delta\ne
1$. However, when the transition takes place between an ordered regime and
a disordered one, the system remains gapless on the ordered side of the
transition point and is gapful on the other side. Therefore, level-crossing
of excited states is absent in this case. In our studies, the low-lying
spectrum of the transverse-field Ising model has this character at
$\lambda=1$.

On the other hand, the concurrence measures actually entanglement between two
localized spins in a mixed state. Therefore, its behavior is also under the
influence of the excited states of the system, especially the low-lying excited
state. To see this point more clearly, let us first consider a generalized
model in one dimension: $ \hat{H} = \sum_i \left(J_x\hat{s}_i^x \hat{s}_{i+1}^x
+ J_y \hat{s}_i^y\hat{s}_{i+1}^y + J_z \hat{s}_i^z \hat{s}_{i+1}^z\right) + h
\hat{s}_i^z$ which can be transformed into the 1D $XXZ$ model and the Ising
model by appropriate choice of the parameters. We then introduce the identity:
$ \langle 0|[A, [H, A]]|0\rangle = 2\sum_n(E_n-E_0)|\langle 0|A|n\rangle|^2$,
where $|n\rangle$ and $E_n$ denote the eigenstate and the corresponding
eigenvalue. The choice of operator $A$ depends on the nature of local ordering.
For the $XXZ$ model near $\Delta=1$, antiferromagnetic order dominates so we
take $A = \hat{s}_\pi^\alpha=\sum_j e^{ij\pi}\hat{s}_j^\alpha$ and obtain
\begin{eqnarray}
-\sum_{\alpha}\langle\hat{s}_j^\alpha\hat{s}_{j+1}^\alpha\rangle
=\frac{E_0}{JN}+\frac{1}{NJ}\sum_{n\alpha} (E_n-E_0)|\langle
0|\hat{s}_\pi^\alpha|n\rangle|^2,\label{eq:xxzconcurrence}
\end{eqnarray}
where $J=2+\Delta$. Obviously, the lhs of Eq. (\ref{eq:xxzconcurrence}) is
directly related to the concurrence of the $XXZ$ model $C=
-2\sum_{\alpha}\langle\hat{s}_j^\alpha\hat{s}_{j+1}^\alpha\rangle-1/2$.
While for the Ising model, the ferromagnetic order dominates so $A =
\hat{s}_0^\alpha =\sum_j \hat{s}_j^\alpha$
\begin{eqnarray}
&& \langle\hat{s}_j^x\hat{s}_{j+1}^x\rangle -
\langle\hat{s}_j^y\hat{s}_{j+1}^y\rangle - \langle\hat{s}_j^z\hat{s}_{j+1}^z\rangle \nonumber \\
&& =-\frac{E_0}{JN}- \frac{1}{NJ}\sum_{\alpha n}(E_n-E_0)|\langle
0|\hat{s}_0^\alpha|n\rangle|^2, \label{eq:isingconcurrence}
\end{eqnarray}
where $J=-\lambda$. The lhs of Eq. (\ref{eq:isingconcurrence}) is also
directly related to the concurrence of the Ising model $C=
2(\langle\hat{s}^x_i\hat{s}^x_j\rangle -
\langle\hat{s}^y_i\hat{s}^y_j\rangle -
\langle\hat{s}^z_i\hat{s}^z_j\rangle)-1/2$. The above two equations tell us
that the concurrence does not simply depend on the re-scaled density of
ground-state energy, but also the contributions from low-lying excited
state with non-zero transition amplitude of the order operator to the
ground state.

Keeping the above facts in mind, we now explain why the concurrence in the
$XXZ$ model reaches its maximum at $\Delta=1$. Notice that, on the
left-hand side of this point, the lowest excited states are doubly
degenerate and have spin numbers $S=1,\>S_z=\pm 1$. Correspondingly, the
matrix elements of spin operators $\hat{s}^+_\pi$ and $\hat{s}^-_\pi$
between them and the singlet ground state $|0\rangle$, are nonzero. On the
other hand, the second excited state has spin number $S=1$ and $S_z=0$ and
it contributes to the longitudinal spin correlation function. On the
right-hand side of the transition point, these excited states interchange
their position, as shown in Fig. \ref{figure_energy}. Since both sides are
ordered phases, the main contributions to the corresponding order
parameters are from the lowest excited state. Then the rhs of Eq.
(\ref{eq:xxzconcurrence}) can be written as $
\frac{E_0}{JN}+\frac{1}{NJ}\sum_{\alpha} (E_1-E_0)|\langle
0|\hat{s}_\pi^\alpha|1\rangle|^2$ approximately. Therefore, the main
contribution to the concurrence is from transverse order operator in the
$XY$ regime and from longitudinal order operator in Ising region. Only at
the transition point $\Delta=1$, all three excited states have the same
energy, and both the transverse and longitudinal spin correlation functions
are power-law decay. Consequently, the contribution from both order
operators to the concurrence makes it maximal.

However, the above argument for the $XXZ$ model is not valid for the
transverse-field Ising model. For the latter, if $\lambda>1$, the ground state
is ferromagnetic and has parity $P=1$, while the lowest excited has parity
$P=-1$. So the rhs of Eq. (\ref{eq:isingconcurrence}) is mostly contributed
from the first excited state and the decreasing of the concurrence as $\lambda$
increases can be well understood. When $\lambda$ approaches the critical point,
the gap formation in the thermodynamic limit will introduce a significant
change to the concurrence. This feature is reflected from the appearance of the
minimum of the concurrence's first derivative at the critical point. On the
other hand, if $\lambda < 1$, the phase is disordered and gapful. In this
situation, though the matrix element of order parameter between the ground
state and the first excited state becomes smaller and smaller, the second
excited state and other higher excited state now can not be neglected. Their
participation not only compensates the lose from the first excited, but also
makes the concurrence to be maximal at one point. However, when
$\lambda\rightarrow 0$, all excited state depart far away from the ground state
which leads to the decrease of the concurrence. Finally, for the Ising model,
its singular behavior around the critical point just results from the
transition from paramagnetic phase to ferromagnetic long-range order phase as
discussed in Refs. \cite{Sachdev,SJGu05}.

With this understanding, let us take another look at the $J_1-J_2$ model.
Besides the first-order transition point $J_2/J_1=0.5$, White and Affleck
found a second-order transition at $J_2/J_1\sim 0.241$ by numerical
calculation \cite{SRWhite96}. Around this point, the ground state of the
model is nondegenerate but a level-crossing between two lowest excited
states occurs as shown in Fig. \ref{figure_energy}. Again, a similar
equation of the concurrence involving matrix element of antiferromagnetic
order parameter can be obtained. Unlike the XXZ model at $\Delta=1$, one of
the excited states involved in level-crossing is a spin singlet and all the
matrix elements of spin operators between it and the ground state, which is
also a spin singlet, are zero. Consequently, this exicted state does not
affect the concurrence at all so the concurrence does not show a maximum
around $J_2/J_1\sim 0.241$. Another excited state with spin $S=1$ involved
in level-crossing is the only one who contributes to the concurrence
significantly. Notice that this state is gapless for $J_2/J_1<0.241$ and is
gapful otherwise. Therefore, the transition is actually of
order-to-disorder type, which we observed in the transverse-field Ising
model. In this case we expect that the critical behavior of the concurrence
will show up in its higher-order derivatives. To check this statement, we
show minimum of higher order derivatives of the concurrence as functions of
system size in Fig. \ref{figure_j1j2dcon}. Although lack of sufficient data
for careful finite size scaling analysis makes it unclear one of the minima
will tend to $J_2=0.241$, we believe that there exists a certain order
derivative of the concurrence whose minimum will tend to 0.241 for an
infinite system. Such behavior is consistent with our picture. On the other
hand, the concurrence is quite flat in the ordered phase and has its
maximum at $J_2/J_1=0$ (Fig. 1). This is due to the fact that the energy
difference between the first excited state and the grounds state is almost
a constant and the antiferromagnetic order parameter has its maximum at
this point.

Finally, for type II and III, we argue that the singularity of the
concurrence at the critical point arises from the change of the long-range
order. First, if the system does not have long-range order, then local
properties of the system, such as energy density, nearest-neighbor
spin-spin correlation, etc., are not affected by spins far away, and their
properties do not depend on the size of the system very much, nor does the
concurrence. On the other hand, if the system has long-range order, the
correlation function in momentum space will be $\delta$-function like,
e.g., $F(g)\delta(q-\pi)$ for antiferromagnetic long-range order, where
$F(g)$ is a coupling dependent function. Then the nearest-neighbor
correlation function is, $\langle 0|\sigma^\alpha_i
\sigma_{i+\delta}^\alpha|0\rangle \propto F(g)$. Thus, the change of
long-range order at the critical point, e.g., from paramagnetic phase to
magnetic order phase for the Ising model \cite{Sachdev}, or from
longitudinal magnetic order to that in the $xy$-plane for the 2D \& 3D XXZ
models\cite{XXZLRO}, will obviously result in the change of $F(g)$ and
leads to a singularity in the correlation function as well as the
concurrence at the critical point. The size-dependent scaling behavior of
the concurrence then comes naturally.

In summary, based on the properties of the low-lying excitation spectrum near
the quantum transition points, we examined and classified the critical
behaviors of the concurrence, a measure of pairwise entanglement, in several
typical spin models. We show that discontinuity of the concurrence in $J_1-J_2$
model is simply caused by the ground-state level crossing at the transition
point. On the other hand, QPT of the $XXZ$ model at $\Delta=1$ is of
ordered-to-ordered type and hence, is accompanied by a level-crossing between
its lowest excited states. Consequently, its concurrence has a maximum at the
transition point. Finally, for the transverse-field Ising model, whose
transition at $\lambda=1$ is of ordered-to-disordered type, the opening of
energy gap at the transition point introduces an extremum to the high-order
derivative of the concurrence. The singularity of the concurrence is the
consequence of the transition of long-range-order. This classification gives a
unified picture on the connection between the QE and QPTs.


This work is supported by RGC Projects CUHK 401703 and 401504 and by CNSF
grant No. 90403003.


\begin{references}

\bibitem{Sachdev}
S. Sachdev, {\it Quantum Phase Transitions}, (Cambridge University
Press, Cambridge, UK, 2000).




\bibitem{Einstein} A. Einstein, B. Podolsky, and N. Rosen,
Phys. Rev. {\bf 47}, 777 (1935).

\bibitem{Schrodinger} E. Schr\"odinger, Naturwissenschaften
{\bf 23}, 807 (1935).


\bibitem{WKWootters98}
W. K. Wootters, Phys. Rev. Lett. {\bf 80}, 2245-2248 (1998); S. Hill and W. K.
Wootters Phys. Rev. Lett. {\bf 78}, 5022-5025 (1997).



\bibitem{AOsterloh2002}
A. Osterloh, Luigi Amico, G. Falci and Rosario Fazio, Nature {\bf
416}, 608 (2002).

\bibitem{TJOsbornee}
T. J. Osborne and M.A. Nielsen, Phys. Rev. A {\bf 66}, 032110(2002).

\bibitem{GVidal2003}
G. Vidal, J. I. Latorre, E. Rico, and A. Kitaev, Phys. Rev. Lett.
{\bf 90}, 227902 (2003); J. I. Latorre, E. Rico, and G. Vidal,
quant-ph/0304098 (2003).

\bibitem{SJGu03}
S. J. Gu, H. Q. Lin, and Y. Q. Li, Phys. Rev. A {\bf 68}, 042330
(2003).

\bibitem{OFSyljuasen03}
Olav F. Sylju{\aa}sen  Phys. Rev. A 68, 060301 (2003).

\bibitem{SJGu05}
S. J. Gu, G. S. Tian, H. Q. Lin, Phys. Rev. A {\bf 71}, 052322 (2005).

\bibitem{JVidal04}
J. Vidal, G. Palacios, and R. Mosseri, Phys. Rev. A {\bf 69}, 022107
(2004).

\bibitem{JVidal042}
J. Vidal, R. Mosseri, J. Dukelsky, Phys. Rev. A 69, 054101 (2004).

\bibitem{LAWu04}
L. A. Wu, M. S. Sarandy, and D. A. Lidar, Phys. Rev. Lett. {\bf 93}, 250404
(2004).

\bibitem{MFYang05}
M. F. Yang, Phys. Rev. A 71, 030302 (2005).


\bibitem{SJGu04J1J2}
S. J. Gu, H. Li,Y. Q. Li, and H. Q. Lin, Phys. Rev. A {\bf 70}, 052302 (2004);
Recep Eryi\u{g}it, Resul Eryi\u{g}it, and Yi\u{g}it G\"{u}nd\"{u}\c{c}, Int. J.
Mod. Phys. C {\bf 15}, 1095 (2004).


\bibitem{CKMajumdar69}
C.K. Majumdar and D.K. Ghosh, J. Math. Phys. {\bf 10}, 1388 (1969).

\bibitem{MANielsenb}
M. A. Nielsen and I. L. Chuang, {\it Quantum Computation and
Quantum information} (Cambridge University Press, Cambridge,
2000).



\bibitem{OFSsen197}
See for example, O. F. Sylju{\aa}sen1, S. Chakravarty, and M. Greven, Phys.
Rev. Lett. {\bf 78}, 4115 (1997).

\bibitem{Lieb} E. Lieb and D. Mattis, J. Math. Phys. {\bf 3}, 749 (1962).

\bibitem{Tian03}
G. S. Tian and H. Q. Lin, Phys. Rev. B {\bf 67}, 245105 (2003).


\bibitem{SRWhite96}
S. R. White and I Affleck, Phys. Rev. B {\bf 54}, 9862 (1996), and the
references therein.


\bibitem{XXZLRO}
F. J. Dyson, E. H. Lieb, and B. Simon, J. Stat. Phys. {\bf 18}, 335 (1978);  T.
Kennedy, E. Lieb, and B. S. Shastry, J. Stat. Phys. {\bf 53}, 1019 (1988).

\end{references}
\end{document}